\documentclass[a4paper]{ceurart}

\usepackage{amsmath,amssymb,amsfonts}
\usepackage{algorithmic}
\usepackage{graphicx}
\usepackage{textcomp}
\usepackage{xcolor}

\usepackage{tikz}
\usetikzlibrary{shapes,arrows}
\usepackage{booktabs}
\usepackage{siunitx}
\usepackage{caption}
\usepackage{subcaption}
\usepackage{orcidlink}
\usepackage{hyperref}

\begin{document}

\copyrightyear{2021}
\copyrightclause{Copyright for this paper by its authors.
  Use permitted under Creative Commons License Attribution 4.0
  International (CC BY 4.0).}

\conference{IPIN 2021 WiP Proceedings, November 29 -- December 2, 2021, Lloret de Mar, Spain}

\title{Multipath-assisted Radio Sensing and Occupancy Detection for Smart In-house Parking in ITS}

\author[1]{Jonas Ninnemann}[%
orcid=0000-0001-7988-079X,
email=Jonas.Ninnemann@tu-dresden.de,
]
\author[1]{Paul Schwarzbach}[%
orcid=0000-0002-1091-782X,
email=Paul.Schwarzbach@tu-dresden.de,
]
\author[1]{Oliver Michler}[%
orcid=0000-0002-8599-5304,
email=Oliver.Michler@tu-dresden.de,
]
\address[1]{Chair of Transport Systems Information Technology, Institute of Traffic Telematics, Technische Universit\"at Dresden, 01062 Dresden, Germany}


\begin{abstract}
Joint, radio-based communication, localization and sensing is a rapidly emerging research field with various application potentials. Greatly benefiting from these capabilities, smart city, mobility, and logistic concepts are key components for maximizing the efficiency of modern transportation systems. In urban environments, both the search for parking space and freight transport are time- and space-consuming and present the bottlenecks for these transportation chains. Providing location information for these heterogeneous requirement profiles (both active and passive localization of objects), can be realized by using retrofittable wireless sensor networks, which are typically only deployed for active localization. An additional passive detection of objects can be achieved by assessing signal reflections and multipath properties of the transmission channel stored within the Channel Impulse Response (CIR). In this work, a proof-of-concept realization and preliminary experimental results of a CIR-based occupancy detection for parking lots are presented. As the time resolution is dependent on available bandwidth, the CIR of Ultra-wideband transceivers are used. For this, the CIR is smoothed and time-variant changes within it are detected by performing a background subtraction. Finally, the reflecting objects are mapped to individual parking lots. The developed method is tested in an in-house parking garage. The work provided is a foundation for passive occupancy detection, whose capabilities can prospectively be enhanced by exploiting additional physical layers, such as 5G or even 6G.
\end{abstract}


\maketitle

\vspace{-0.5cm}
\section{Introduction}
\label{sec:Intro}
Intelligent transportation systems profit from the interconnection and localization of vehicles, objects, and traffic participants. In this context, location-aware communication is an essential integral part for increasing the efficiency, safety and environmental friendliness of transportation and supply chains. With the rapidly increasing urbanization and growing logistical volumes, intelligent and future-oriented mobility solutions, based on the advancing digitization and technological progress, have to be provided. For this task, concepts for smart cities are studied and discussed \cite{Kirimtat2020Smart_Cities}, where mobility demands are a major challenge. Especially for motorized individual traffic in urban environments, the search for parking space is time-consuming, thus hindering the capabilities of efficient traffic flow management. 

In this context, smart parking systems \cite{Lin2017Smart_Parking_Survey} provide sensor-based and networked solutions to detect parking lot occupancy rates and dispense available capacities in order to minimize parking lot searches. While online smart parking systems offer great potential for compensating the aforementioned issues of urban parking \cite{Martino2019Parking_Guidance_Advantages, Rabby2019Smart_traffic_management}, sensory perception of the real-time occupancy rate of available parking space is essential for the performance of smart parking systems. For this task, manifold approaches for providing efficient and reliable occupancy detection of parking space exist, including visual sensor networks with camera via image recognition \cite{Baroffio2015Visual_SN_Parking_lot_detection, Ho2019Computer_vision_occupation_system}, communication-based via vehicular ad hoc networks \cite{Caliskan2007VANET_Occupancy}, crowd sensing by taxis and map/location information \cite{Bock2020Taxi_Crowd} or camera-based drone surveillance \cite{Sarkar2019Drone_surveillance}. More conventional systems typically require parking space selective infrastructure, such as magnetic or infrared \cite{Paidi2018smartparkingsensors}, or more recently radio-based sensors \cite{Solic2019Wireless_Presence_Detection}. 

Additionally, Wireless Sensor Networks (WSN) provide cheap, long-lasting, and efficient technological solutions for various applications \cite{Zafari2019IPSSurvey}, based on the interconnection of devices, while also providing active localization capabilities. Recently, radio-based localization and networking systems focusing on indoor parking have been developed \cite{icinco20}, in order to provide cheap and retrofittable solutions for vehicular guidance, occupancy rate and digital billing. 

Concurrently to increasing demands on parking space distribution, concepts of further using parking space for logistic applications have been formulated \cite{Faugere2020Sustainability_logistic_hubs}. These provide the idea of additionally using in-house parking lots as so-called mini-hubs for logistical applications to further increase the efficiency of spatial resources by occupying parking space with freight trailers or other storage capabilities.

While this approach leads to increasing capacity optimization, it also changes the demands for localization systems as a basis for occupancy state detection. For the detection of vehicles, active transponders (either the connected vehicle itself, a smartphone, or a connected device at the entrance) are typically required. However, freight trailers or other storage objects are not equipped with these. 
This hybridization of both active and passive objects within the environment requires additional surveillance capabilities. 

In this context, the recently emerging field of joint communication and sensing \cite{zhang2021JCAS, 6G} provides technologically advantageous approaches for this task, as it allows the re-use of radio signals for radar-like applications, e.g.\ bistatic or multistatic radars \cite{kanhere2021target}, and environmental sensing \cite{barneto2021radiobased}, both of which is also referred to as device-free passive localization \cite{jovanoska_device-free_2013-1}. 

\begin{figure}[pos=ht]
    \centering
    \includegraphics[width=0.67\linewidth]{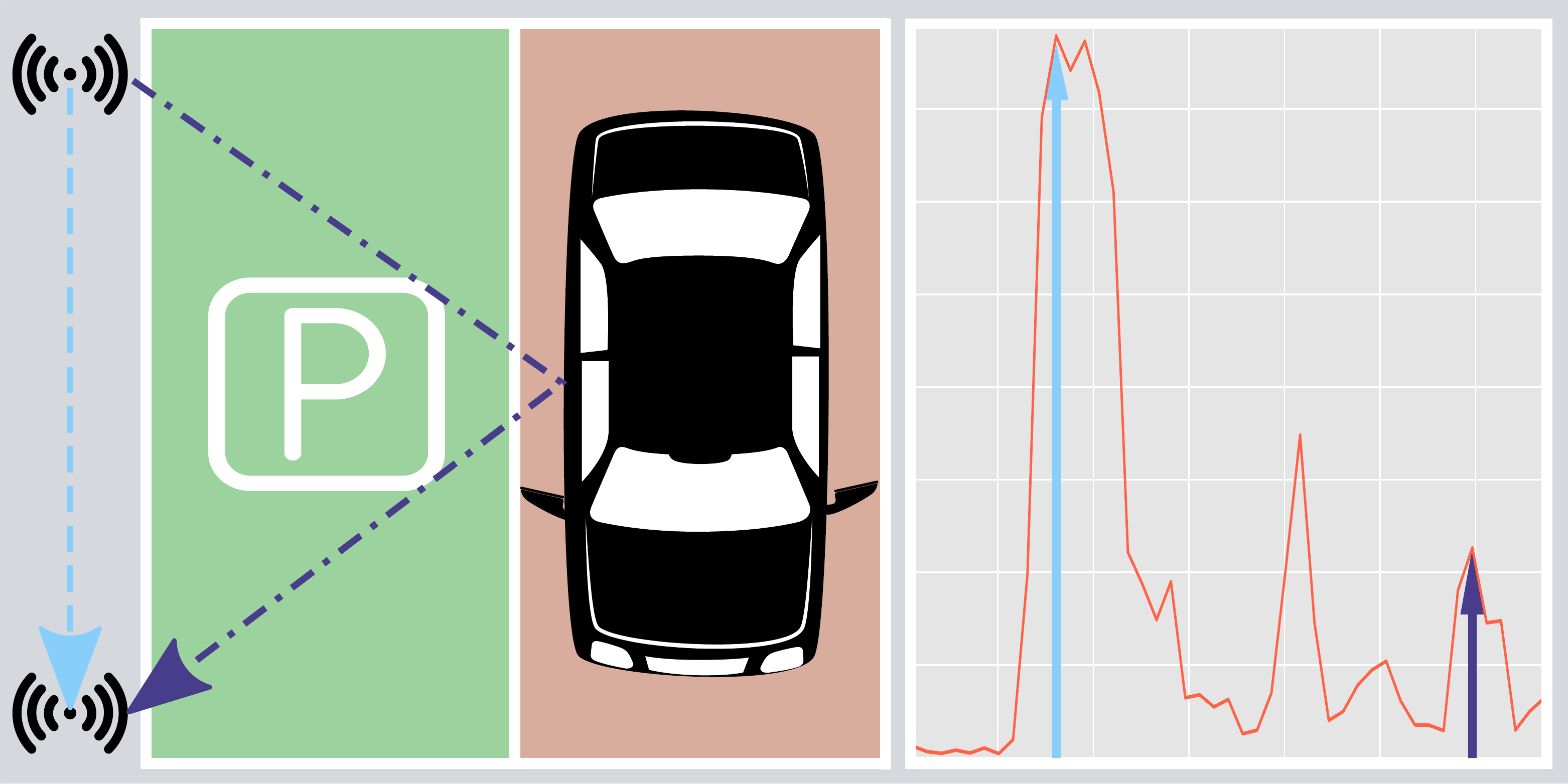}
    \caption{Joint communication, localization, and sensing for in-house parking lots: Radio-based passive occupancy detection enabled by mapping the environmentally dependent multipath propagation using the Channel Impulse Response (CIR). The direct and an exemplary reflection path are depicted in light blue and blue in both the site plan (left) and the CIR (right). Free and occupied parking lots are indicated in green and red.}
    \label{fig:Multipath}
\end{figure}

This paper focuses on the merging of an already established active localization framework for in-house parking based on Ultra-wideband (UWB), which was presented in \cite{icinco20}, and the consecutive use of the transmitted signals by analyzing the Channel Impulse Response (CIR) of the transmission channel. The information on multipath propagation stored in the CIR indicate the presence of reflecting objects, in this special case parking vehicles or trailers, which allow conclusions on the occupancy state of the area (cf. Fig.~\ref{fig:Multipath}). Furthermore, the UWB CIR can also be used for passive localization of reflecting objects \cite{sensors_DFPL, Cimdins2020MAMPI_UWB}.

In order to detect the occupancy of parking space, we use an exponentially weighted moving average (EWMA) filter to smooth the CIR and then perform a background subtraction to detect differences between a static and non-occupied CIR and a potentially occupied measurement input. Subsequently, the detected changes within the CIR are mapped to the environment using an elliptical model and a heatmap representation. 

\vspace{0.5 cm}

The developed approach is empirically examined in a demanding real-world, in-house parking scenario using a set of UWB transceivers and a minibus, which could potentially represent both parking vehicles and freight trailers. In this context, this paper focuses on the recognizability of the occupation solely derived from the measured CIR and discusses the challenges and limitations of different constellations as well as the examined UWB physical layer.  

The rest of the paper is structured as follows: After the introduction in Sec.~\ref{sec:Intro}, Sec.~\ref{sec:Approach} presents the passive, radio-based occupancy detection approach as well as necessary inputs and processing steps. Subsequently, Sec.~\ref{sec:Measurements} introduce the in-house parking measurement environment and the deployed UWB sensors. The surveyed measurements are presented and discussed in Sec.~\ref{sec:Results}. The paper concludes with a summary and proposals for future work in Sec.~\ref{sec:Conclusion}.

\section{Multipath-assisted Passive Occupancy Detection}
\label{sec:Approach}

This section presents the proposed passive occupancy detection approach, for which all processing steps are depicted in Fig.~\ref{fig:Processing}. These will be briefly discussed in the following.

\tikzstyle{block} = [rectangle, draw, fill=gray!20, 
    text width=4.5em, text centered, rounded corners, minimum height=4em]
\tikzstyle{line} = [draw, -latex']

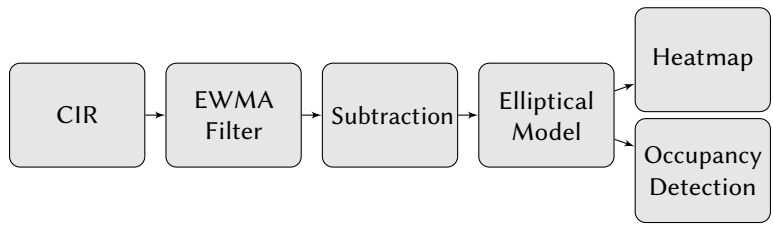
\begin{figure}[pos=ht]
    \centering
    \begin{tikzpicture}[node distance = 2.1 cm]
        \begin{scope}[scale=0.98, transform shape]
            \node [block] (cir) {CIR};
            \node [block, right of=cir] (filter) {EWMA Filter};
            \node [block, right of=filter] (sub) {Subtraction};
            \node [block, right of=sub] (ell) {Elliptical Model};
            \node [block, right of=ell, yshift=-.75cm] (detection) {Occupancy Detection};
            \node [block, right of=ell, yshift=.75cm] (map) {Heatmap};
            \path [line] (cir) -- (filter);
            \path [line] (filter) -- (sub);
            \path [line] (sub) -- (ell);
            \path [line] (ell) -- (detection);
            \path [line] (ell) -- (map);
        \end{scope}
    \end{tikzpicture}
    \caption{Flowchart of multipath-assisted occupancy detection.}
    \label{fig:Processing}
\end{figure}

\subsection{Multipath Propagation}
The foundation for the proposed passive occupancy detection approach is the multipath propagation phenomenon, where a wireless radio signals reaches the receiver's antenna via multiple paths. 
In the past, multipath propagation was tried to be mitigated in order to improve the quality of the communication or, for localization tasks, the distance measurements between sensors. However, additional information can be derived by assessing the transfer function of the transmission channel, as reflections are caused by static or dynamic objects within the propagation environment. This leads to the time-delayed reception of additional signal components. Measuring and identifying multipath propagation can be achieved by using large bandwidths. Currently, UWB is a reasonable technological candidate for this task, as pulses with a very short duration are used. The measured multipath components (MPC) are stored within the CIR.

\begin{figure}[pos=ht]
    \centering
    \includegraphics[width=0.6\linewidth]{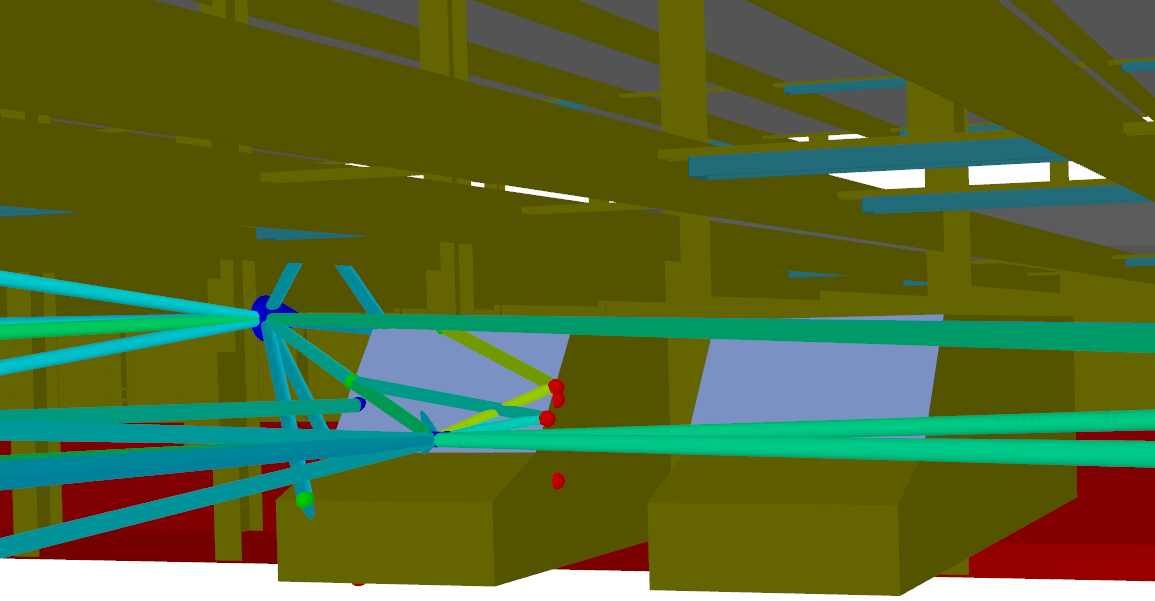}
    \caption{3D Ray-tracing simulation for in-house parking multipath propagation including the transmitting antenna (blue), reflection areas (red dots) and propagation paths (green).}
    \label{fig:funkplanung}
\end{figure}

For heavy-multipath environments, such as an in-house parking area, multiple dominant reflection sources are present, e.g. parking vehicles or concrete. A typical multipath propagation in this scenario is given in Fig.~\ref{fig:funkplanung}, which shows the multipath richness of this environment.

\subsection{Channel Impulse Response}
\label{subsec:Channel}
The CIR represents the time delays caused by the multipath propagation of radio signals. In general, the output $x(t)$ of a wireless communication system in form of the CIR $h(t)$ with respect to the input signal $s(t)$ at time $t$ can be described as

\begin{equation}
x(t)= h(t) \ast s(t) + v(t) = \int h(\tau)s(t-\tau)d\tau+v(t),
\label{equ:LTI}
\end{equation}

where $v(t)$ is the additive noise of the signal \cite{jiang_multipath_2011-1}. Given an impulse input signal, the CIR consists of complex numbers representing the in-phase and quadrature components of the received radio signals. Each value $k$ is a specific time-shifted impulse $\delta(t -\tau_k)$, with $\delta(\cdot)$ representing the delta function. The CIR is defined in (\ref{equ:CIR}), where $\alpha_k$ and $\tau_k$ denote the impulse amplitudes and reception time delays~\cite{matthews_understanding_2019-2}.

\begin{equation}
h(t) = \sum_{k=1}^{K} \alpha_k \delta (t-\tau_k)
\label{equ:CIR}
\end{equation}

Furthermore, the magnitude of the CIR can be calculated from the obtained complex CIR raw data.

\subsection{Exponential Weighted Moving Average Filter}
In order to reduce the influence of outliers, the CIR is filtered over multiple measurement epochs by applying an EWMA filter. Additionally, EWMA filter can be used to extract static backgrounds in dynamic scenarios \cite{klemm_people_2017}.

The averaged CIR $z_t$ at time $t$ is computed using the previous one $z_{t-1}$ and the newly received CIR $h_t$. The parameter $\alpha$ in (\ref{equ:EWMA}) represents a constant scalar weighting factor \cite{jovanoska_multiple_nodate}, which for the described use case was set to $\alpha = 0.3$. Due to the static environment the $n=300$ CIR measurements per scenario are averaged with a small window size of $k=5$ by the EWMA filter.

\begin{equation}
z_t = \alpha h_t + (1-\alpha)z_{t-1}
\label{equ:EWMA}
\end{equation}

\begin{figure}[pos=ht]
    \centering
    \includegraphics[width=0.6\linewidth]{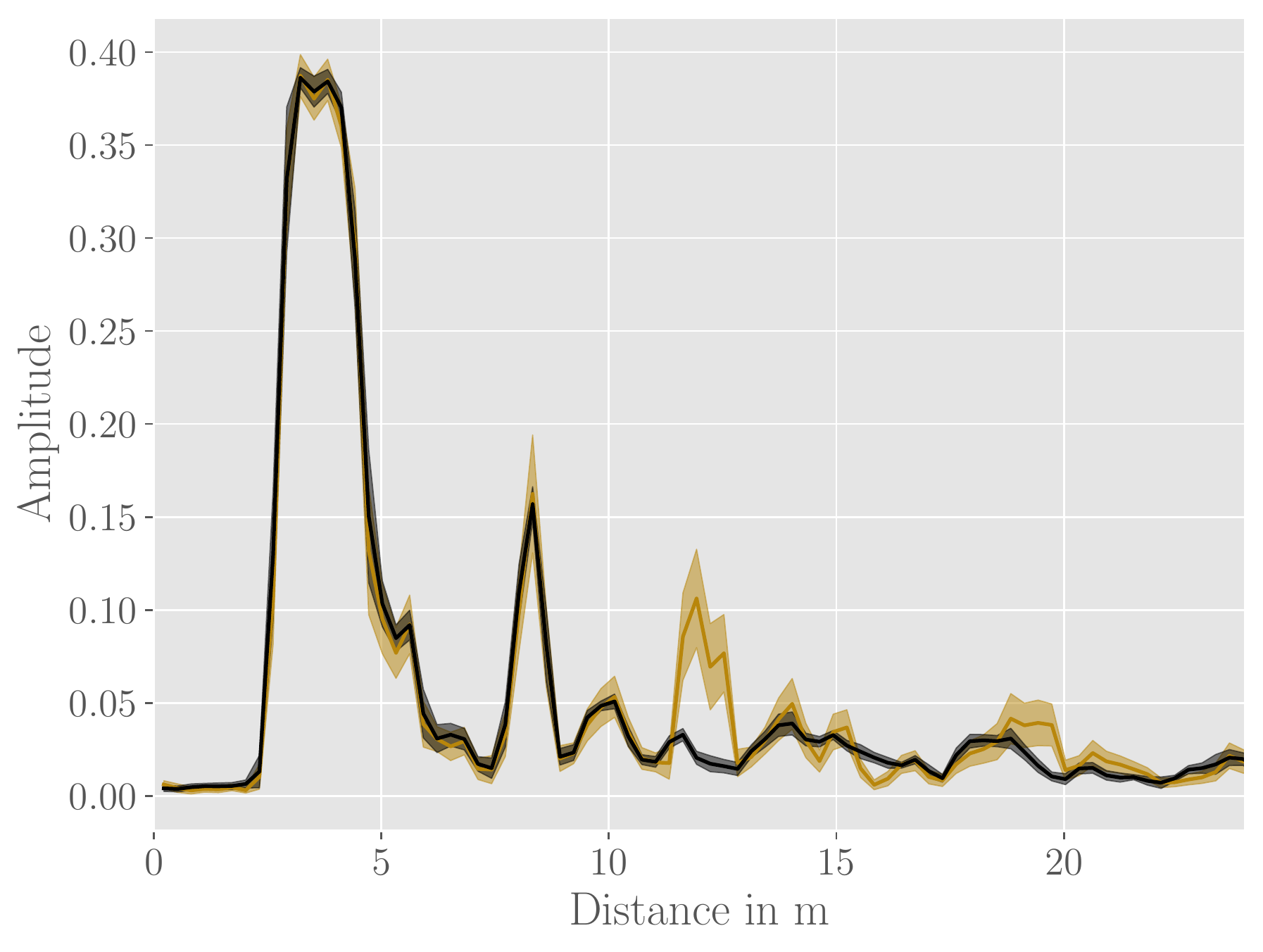}
    \caption{EWMA filtered CIR with (golden) and without (black) vehicle ($n=300$).}
    \label{fig:EWMA}
\end{figure}

Fig.~\ref{fig:EWMA} depicts EMWA results for two different scenarios, including the standard deviation of the filtered results.

\subsection{Background Subtraction}
In complex, real-world scenarios an identification of reflecting sources can be spatially ambiguous if no prior information is available. For this reason, we follow a background subtraction $s_t$ approach for each measurement epoch $t$, which aims at removing static components of the compared CIRs \cite{klemm_people_2017, background_subtraction}. This can be achieved by comparing the filtered CIR $z_t^1$ with a previously surveyed reference CIR $z_t^0$, which represents a non-occupied scene. 

\begin{equation}
s_t = \lvert z_t^1 - z_t^0 \rvert
\label{equ:Subtraction}
\end{equation}

\begin{figure}[pos=ht]
    \centering
    \includegraphics[width=0.6\linewidth]{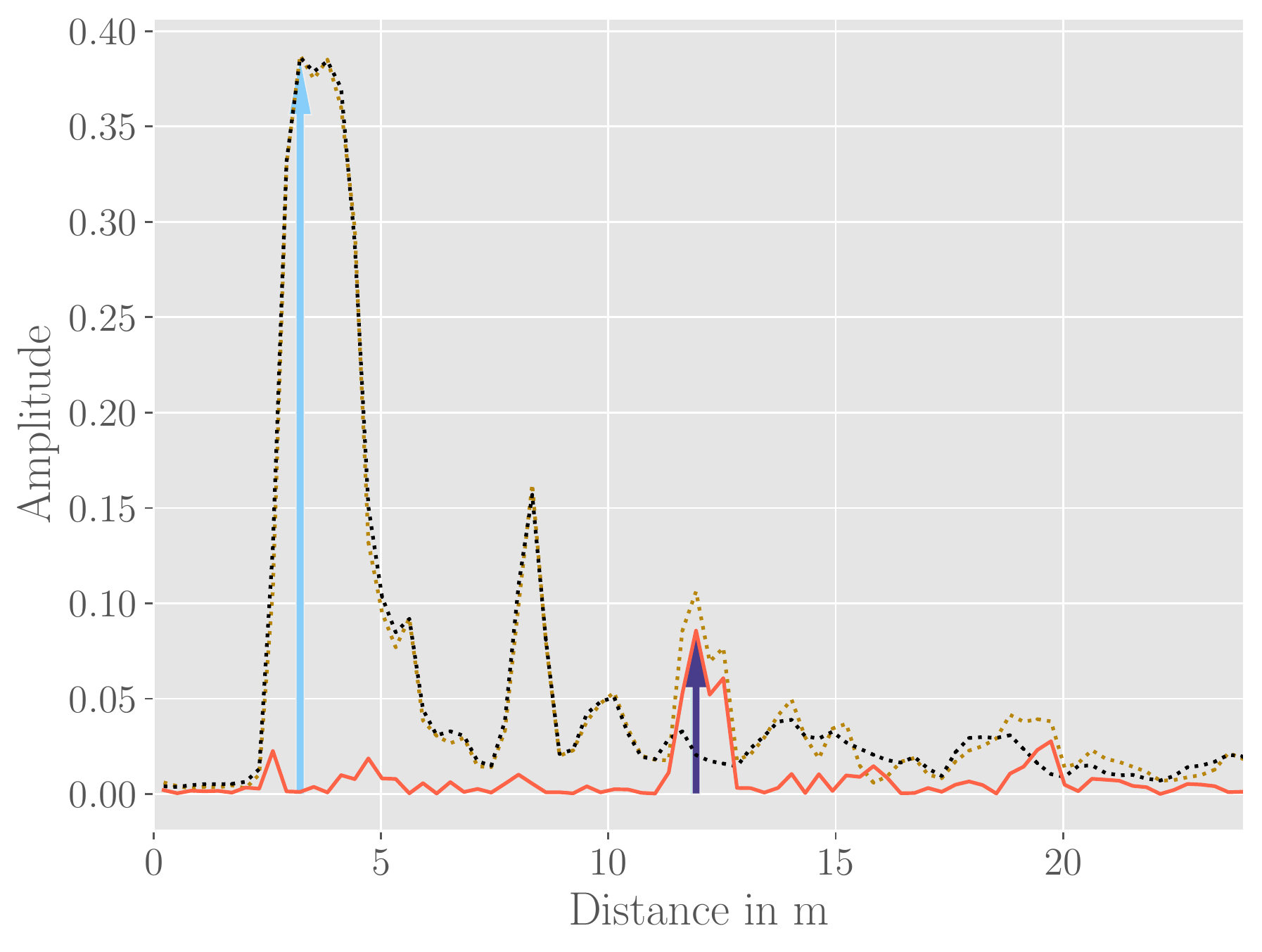}
    \caption{Subtracted CIR (red) with $z_t^1$ (golden) and without vehicle $z_t^0$ (black), including the direct path (light blue) and the estimated reflection path (blue).}
    \label{fig:Subtraction}
\end{figure}

In order to provide a correct overlaying of the compared CIRs, they are synchronized via the direct path and the corresponding index. Therefore, they are synced at the leading edge, which is located in the flank of the peak of the direct signal path. In addition, the amplitude values of the filtered CIRs are normalized for a better comparability. Fig.~\ref{fig:Subtraction} illustrates the subtraction of the CIRs, where both the direct path and an estimation of the reflection path at a target object are also marked. The estimation of the reflection path is obtained as the maximum of $s_t$.
 
\subsection{Elliptical Model and Mapping}
\begin{figure}[pos=ht]
	\centering
    \begin{subfigure}[b]{0.45\textwidth}
        \centering
        \includegraphics[trim=0 10 0 5, clip,width=1\linewidth]{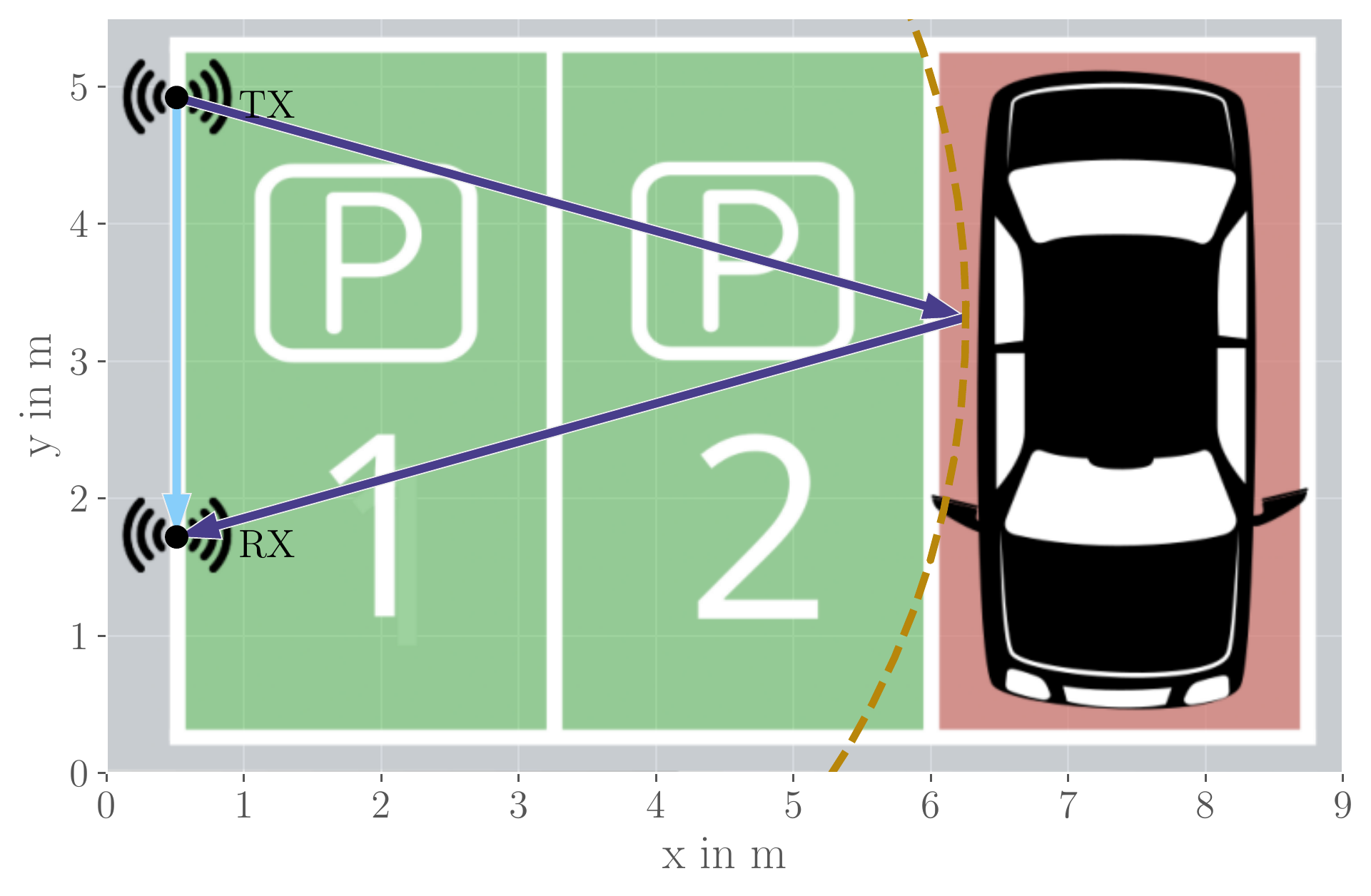}
        \caption{}
        \label{fig:Map}
	\end{subfigure}
	\centering
	\begin{subfigure}[b]{0.53\textwidth}
        \centering
        \includegraphics[trim=0 10 7 5, clip, width=1\linewidth]{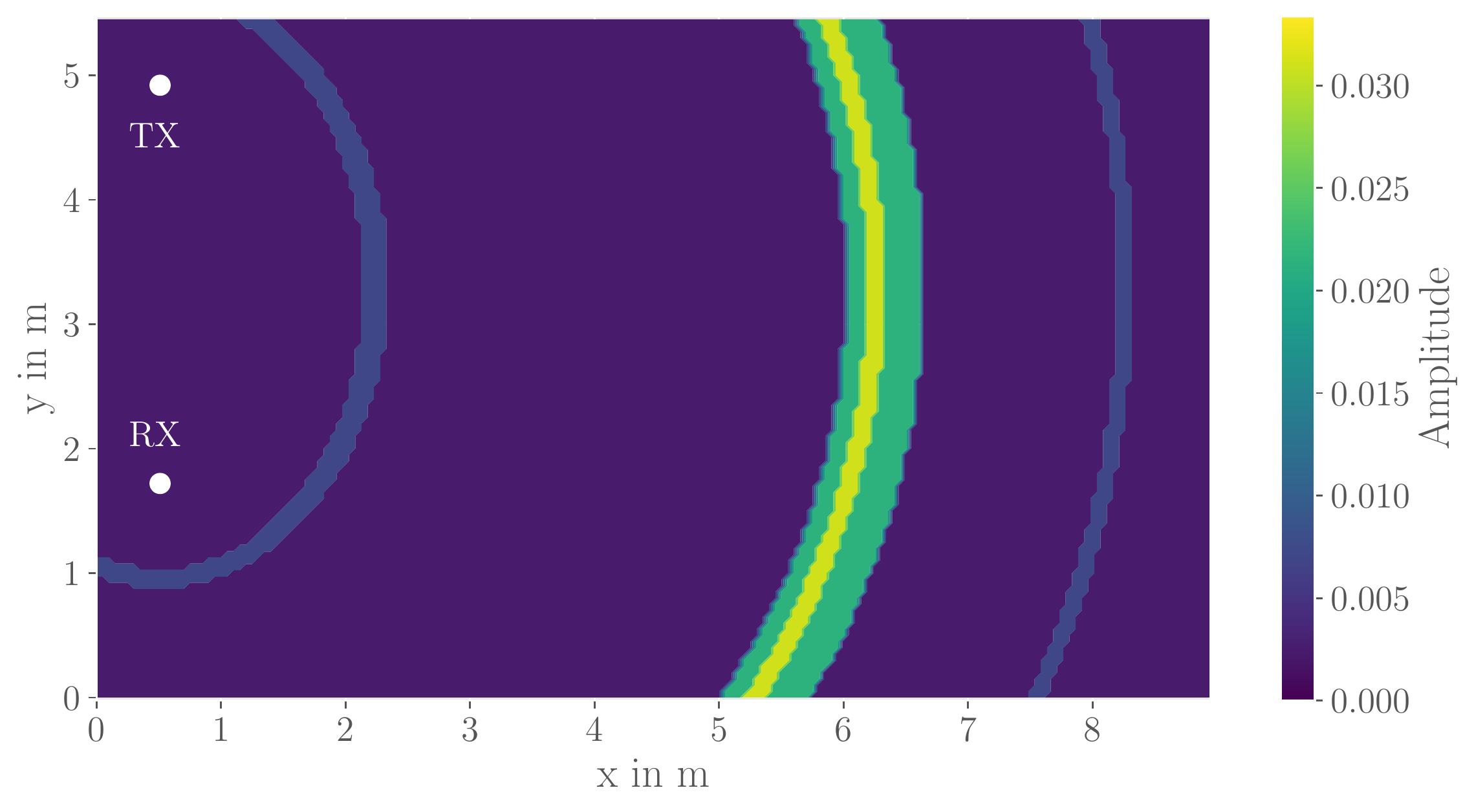}
        \caption{}
        \label{fig:Heatmap}
	\end{subfigure}
	\caption{CIR mapping: \textbf{(a)} Mapping of the reflection path (blue); \textbf{(b)} Heatmap of the environment.}
	\label{fig:sideways_far}
\end{figure}

For the detection of the target vehicle, the time/distance information from the CIR are mapped into a plane. Fig.~\ref{fig:sideways_far} depicts a graphical overview of the corresponding processing steps. The mapping is realized with an elliptical model, which geometrically represents the positional line of the reflection source, given a bistatic setup and the measured path length from the subtracted CIR $s_t$. The position of the transmitter $\textbf{X}_\text{T}~=~(x_{\text{T}},y_{\text{T}})^\intercal$ and the receiver $\textbf{X}_{\text{R}}~=~(x_{\text{R}},y_{\text{R}})^\intercal$ within the communication network have to be known in order to detect the target vehicle $\textbf{X}~=~(x, y)^\intercal$. Eq.~(\ref{equ:path}) represents the length of the reflection path or the bistatic range $d_\text{r}$. 

\begin{equation}
    d_r= \left \|\boldsymbol{X}_\text{T} - \boldsymbol{X} \right \|_2 + 
    \left \| \boldsymbol{X}_\text{R} - \boldsymbol{X} \right \|_2
    \label{equ:path}
\end{equation}

The bistatic range $d_\text{r}$ is calculated from the CIR via the corresponding time delay $\tau$ in the CIR for the reflection path and the speed of light $c$: 

\begin{equation}
d_r = \tau \cdot c.
\label{equ:path2}
\end{equation}

Eq.~(\ref{equ:path}) describes an ellipse with the transmitter and the receiver located in the foci. The semi-major axis $a$ is calculated with the help of the measured bistatic range $d_\text{r}$ and the semi-minor axis $b$ with the direct path length $d_\text{p}$.

\begin{equation}
a = \frac{d_r}{2} \quad \quad \quad \quad b = \sqrt{a^2-(\frac{d_p}{2})^2}
\label{equ:ellipse}
\end{equation}

The two estimated axis of the ellipse allow setting up the parametric equation of the ellipse and therefore the mapping of the reflection path, which is qualitatively shown in Fig.~\ref{fig:Map}.

As aforementioned, every individual values stored within the CIR represents a potential reflection source, hence all CIR values with their corresponding time delay and amplitude are represented as ellipses with the same foci but different semi-major axis. 

In addition, the family of ellipses can be used to provide a representation of the propagation environment and its reflecting sources by performing an environmental mapping. In \cite{sensors_DFPL}, we proposed the creation of a heatmap based on the time delays and amplitudes of each CIR value. The resulting continuous value surface of the heatmap is obtained by performing a nearest neighbor interpolation and shown in Fig.~\ref{fig:Heatmap}.
Given the example in Fig.~\ref{fig:sideways_far}, a spatial identification of the target vehicle can be achieved. With multiple CIR measurements between different anchors an unambiguous localization of the reflecting object is also possible. \cite{sensors_DFPL}

\subsection{Occupancy Detection}
The occupancy detection is performed on the basis of the estimated reflection path from the subtracted CIR. For this purpose, an interval of the reflection path length with respect to the sensor arrangement is calculated for each parking lot. Afterwards the estimated reflection path length and the intervals are compared in order to assign them to a parking lot. For example an estimated reflection path length between \SI{11}{\meter} and \SI{17}{\meter} indicates a vehicle on parking lot P3. For the empirical evaluation, the occupancy detection state is estimated over all measurement epochs and expressed as the number of correct assignments.
\section{In-house Parking Measurements}
\label{sec:Measurements}
\begin{figure}[pos=h]
    \centering
    \begin{subfigure}[b]{0.55\textwidth}
        \centering
        \includegraphics[width=\textwidth]{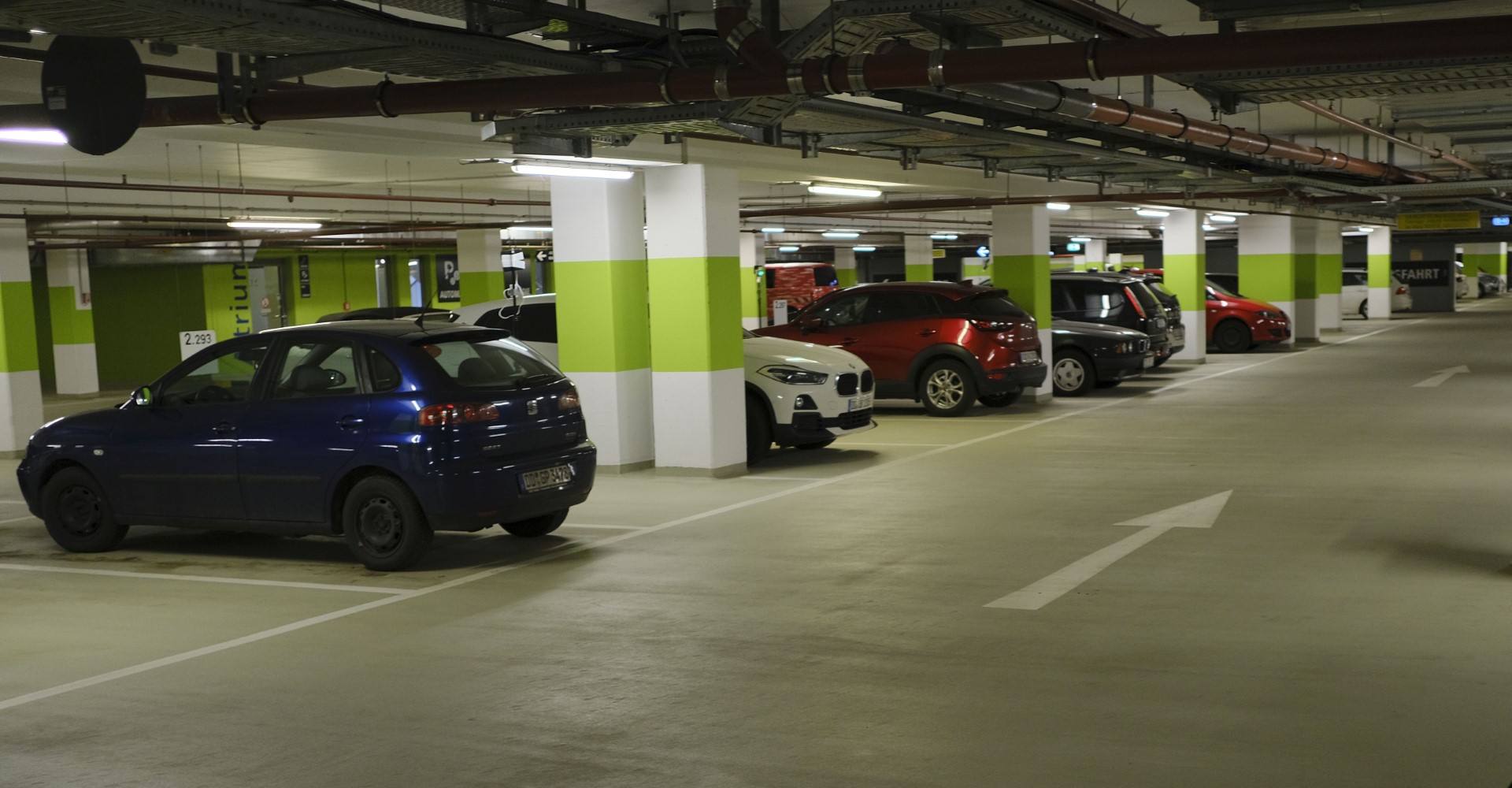}
        \caption{}
        \label{fig:WTC}
	\end{subfigure}
    \centering
    \hspace{0.1cm}
    \begin{subfigure}[b]{0.195\textwidth}
        \centering
        \includegraphics[width=\linewidth]{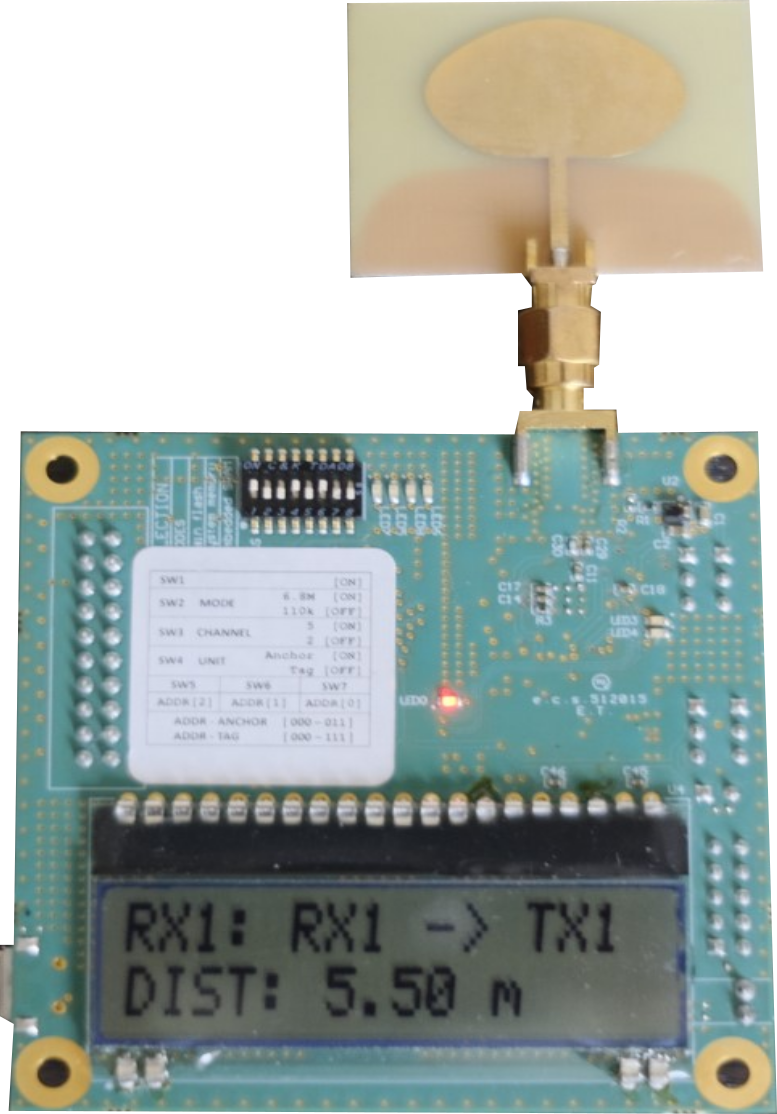}
        \caption{}
        \label{fig:EVB1000}
    \end{subfigure}
    \label{fig:setup}
    \caption{Measurement setup: \textbf{(a)} Parking garage; \textbf{(b)} \textit{Decawave} EVB1000 UWB Transceiver.}
\end{figure}

The conducted measurements were carried out in an in-house parking garage, representing a challenging environment for wireless-enabled localization systems. Based on an already established active UWB-based positioning system, the developed approach can further enable occupancy detection and therefore benefit smart parking systems. The testbed for this is an underground parking garage located in Dresden, Germany with over 1000 available parking spaces on two floors \cite{WTC} (cf. Fig.~\ref{fig:WTC}). Each parking lot is \SI{2.75}{\meter} wide and \SI{5.0}{\meter} long. The target object for the occupancy detection is a minibus with a size of \SI{5.3}{\meter} x \SI{2}{\meter} x \SI{2}{\meter}. 

The CIR measurements are carried out with \textit{Decawave} EVB1000 Transceivers (cf. Fig.~\ref{fig:EVB1000}), equipped with a micro-controller and the \textit{Decawave} DW1000  radio chip. For the conducted measurements the parameters in Tab.~\ref{tab:Parameters} were applied.

\begin{table*}[pos=h]
    \caption{Measurement Parameters of the DW1000.}
	\label{tab:Parameters}
	{\renewcommand{\arraystretch}{1.0}
		\begin{tabular}{p{4cm} p{3cm}}
			\toprule
			Parameter & Value \\
			\midrule
			Center frequency & \SI{4492.8}{\mega \hertz}\\
			Bandwidth & \SI{499.2}{\mega \hertz}\\
			Pulse repetition frequency (PRF) & \SI{16}{\mega \hertz}\\
			Preamble length & 1024\\
			Data rate & \SI{850}{kbps}\\
			CIR time resolution $\Delta t \approx$ & \SI{1}{\nano \second} or \SI{30}{\centi \meter}\\
			\bottomrule
		\end{tabular}
	}
\end{table*}

The UWB transceivers were attached to tripods at a height of \SI{1.5}{\meter} and connected to a laptop. For this preliminary measurement campaign, two transceivers are deployed to record the CIR raw data, which consists of 992 complex, evenly separated values. With regard to the available bandwidth, the time resolution amounts to $\Delta t \approx$ \SI{1}{\nano \second} or \SI{30}{\centi \meter}.

\section{Results}
\label{sec:Results}

For the empirical validation of the proposed method in a demanding real-world scenario, an in-house parking environment and a minibus was chosen, in order to provide a proof-of-concept for passive vehicle or freight trailer detection based on the CIR. Therefore, the focus of the investigated scenarios is on three parking lots and the correct assignment of the occupancy detection state. In addition, the effect of the EWMA filter is validated. The parking lots are numbered with P1, P2, and P3 to distinguish them in the scenarios.

\begin{figure}[pos=h]
    \centering
    \hspace{0.3cm}
	\begin{subfigure}[b]{0.38\textwidth}
		\centering
		\includegraphics[width=1\linewidth]{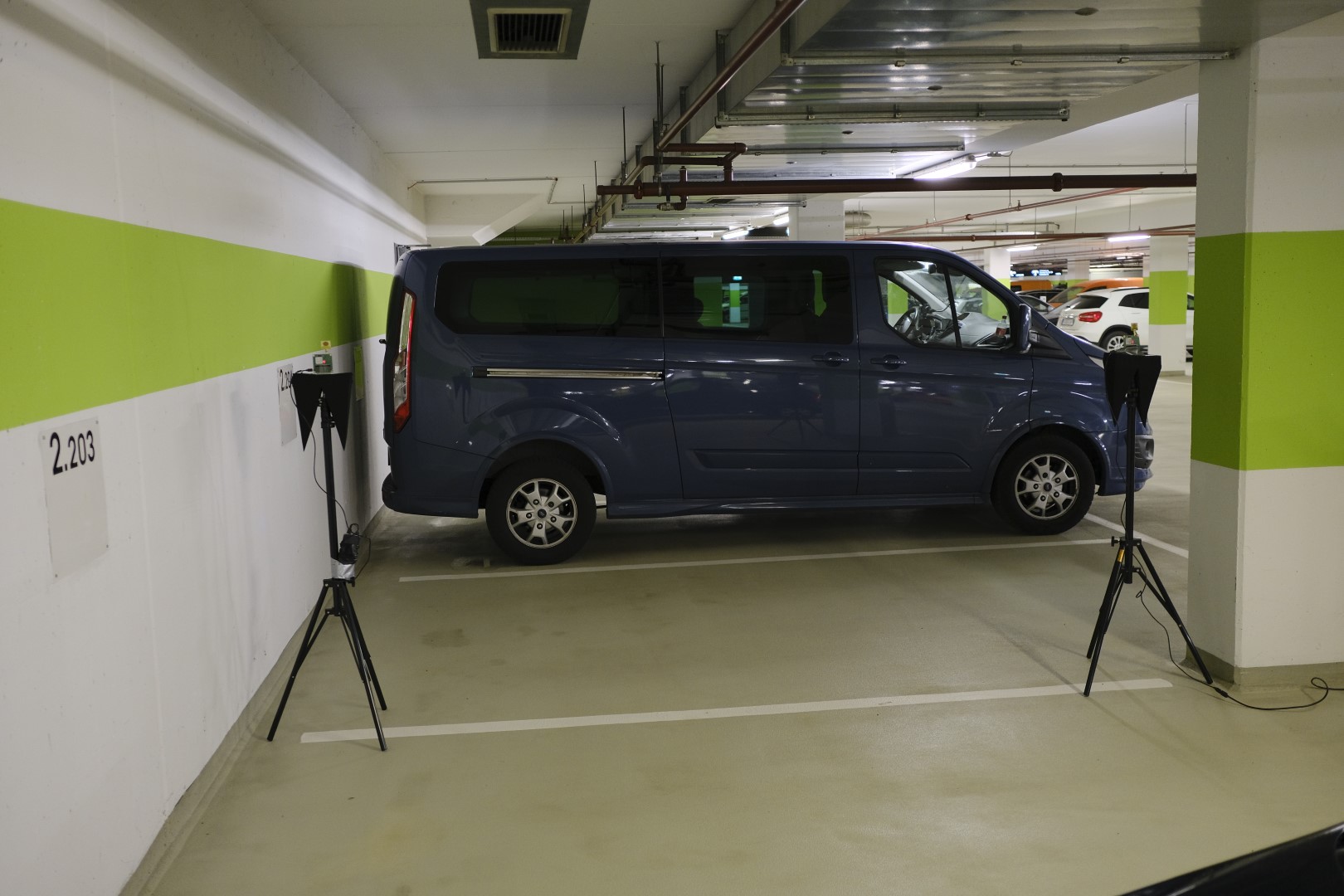}
		\vspace{0.14cm}
		\caption{}
		\label{fig:1_Foto_with}
	\end{subfigure}%
	\hspace{0.6cm}
	\centering
	\begin{subfigure}[b]{0.45\textwidth}
		\centering
		\includegraphics[trim=0 0 0 0, clip, width=1\linewidth]{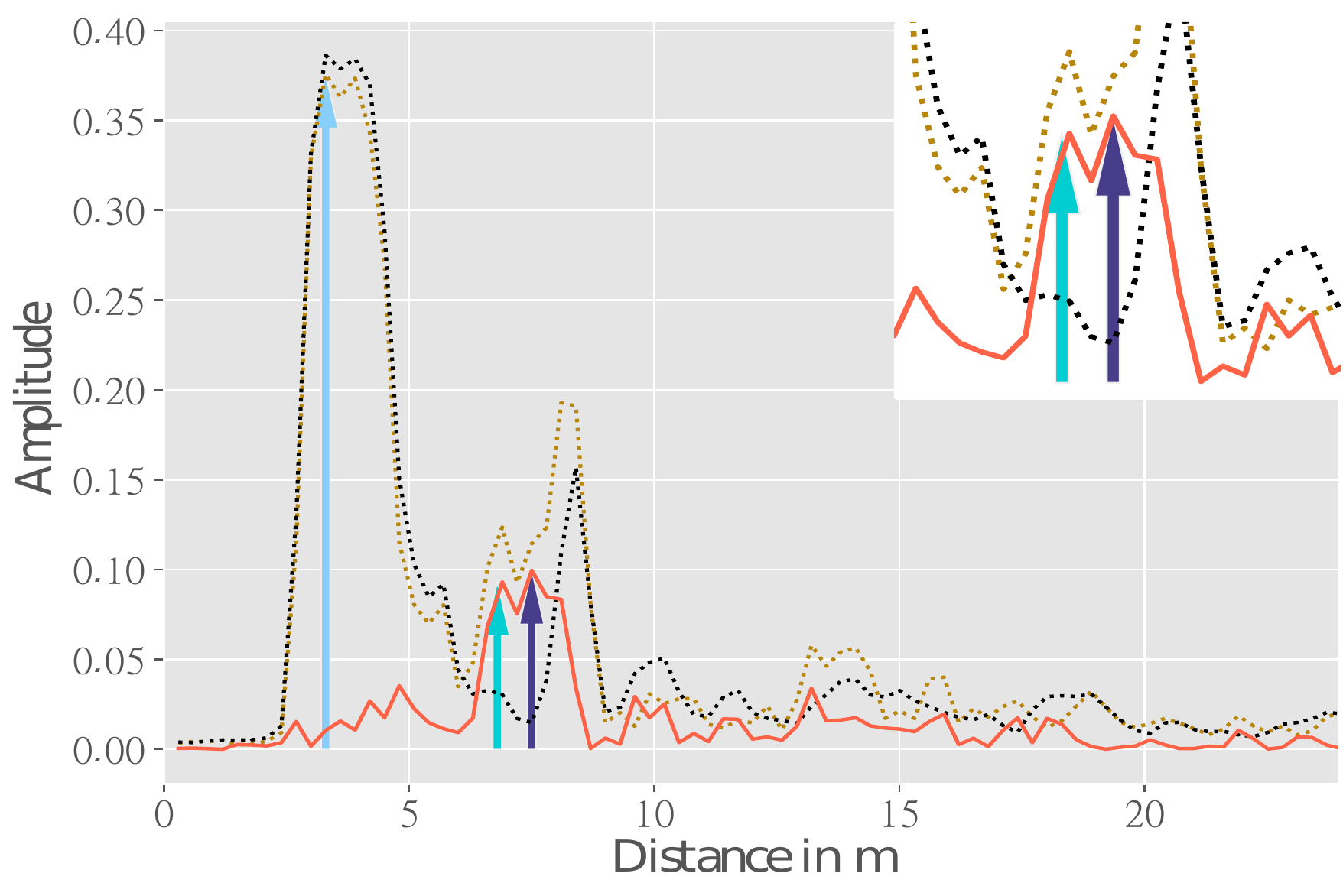}
		\caption{}
		\label{fig:1_CIR}
	\end{subfigure}
	\vspace{0.2cm}
	\begin{subfigure}[b]{0.47\textwidth}
		\centering
		\includegraphics[width=1\linewidth]{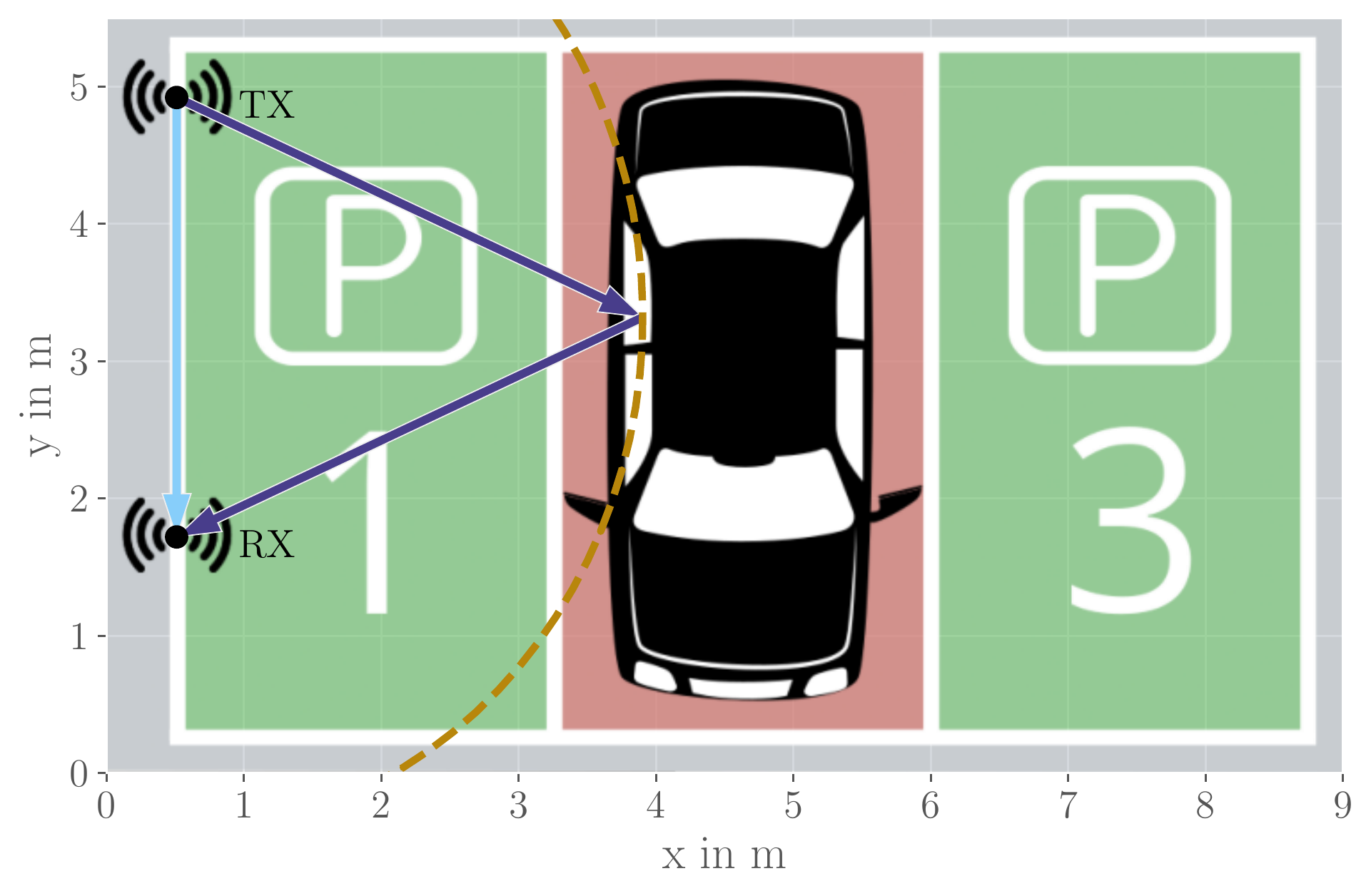}
		\caption{}
		\label{fig:1_Lageplan}
	\end{subfigure}
	\centering
	\hspace{0.1cm}
	\begin{subfigure}[b]{0.47\textwidth}
		\centering
		\includegraphics[trim=0 0 110 0, clip, width=1\linewidth]{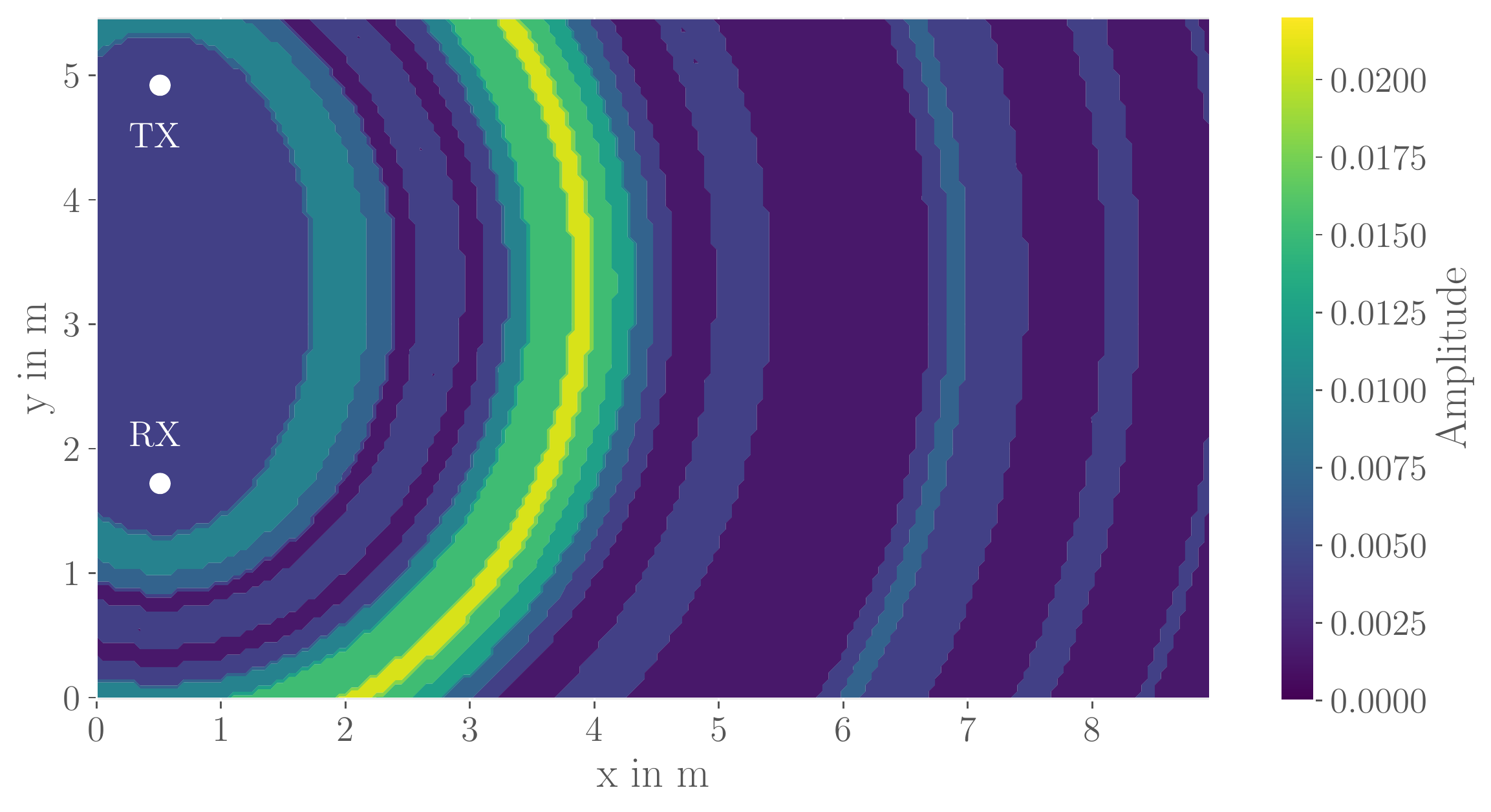}
		\caption{}
		\label{fig:1_Heatmap}
	\end{subfigure}
	\caption{Occupancy detection for parking lot P2: \textbf{(a)} CIRs with subtraction with reference reflection path (cyan); \textbf{(b)} Environment with the minibus; \textbf{(c)} Site plan with estimated reflection path (blue); \textbf{(d)} Interpolated Heatmap.}
	\label{fig:sideways_near}
\end{figure}

In the first configuration, the transceivers are placed on the side of the vehicle, as shown in Fig.~\ref{fig:1_Lageplan}. The transmitter and the receiver are \SI{3.2}{\meter} apart and located at the edge of P1. Three different measurement runs were carried out with this arrangement:

\begin{enumerate}
    \item Empty parking lots for CIR calibration ($z_{t}^0$), 
    \item with a minibus on P2 (cf. Fig.~\ref{fig:sideways_near}) and
    \item with a minibus on P3 (cf. Fig.~\ref{fig:sideways_far}).
\end{enumerate}

\subsection{Minibus on P2}
The results of the detection steps for this scenario are shown in Fig.~\ref{fig:sideways_near}. Based on the geometric arrangement the reference reflection path length at the vehicle is \SI{6.8}{\meter}. 

For the estimation of the reflection path length from the CIR, the EMWA filter and the background subtraction are applied. Here the subtraction is capable of removing other reflections from the environment in order to focus only on the variant reflections when compared to $z_t^0$. The estimated reflection path is \SI{7.5}{\meter} long, which allows a clear mapping and detection of the vehicle. The deviation to the reference reflection length is potentially caused by multiple reflections at the vehicle, as the subtracted CIR in Fig.~\ref{fig:1_CIR} indicates. Also, the CIR uncovers a second reflection with only a slightly larger path length for both measurements. With help of the subtraction it is possible to remove this MPC from the CIR.

\begin{figure}[pos=ht]
	\centering
	\begin{subfigure}[b]{0.49\textwidth}
		\centering
    	\includegraphics[trim=-15 0 0 0, clip, width=1\linewidth]{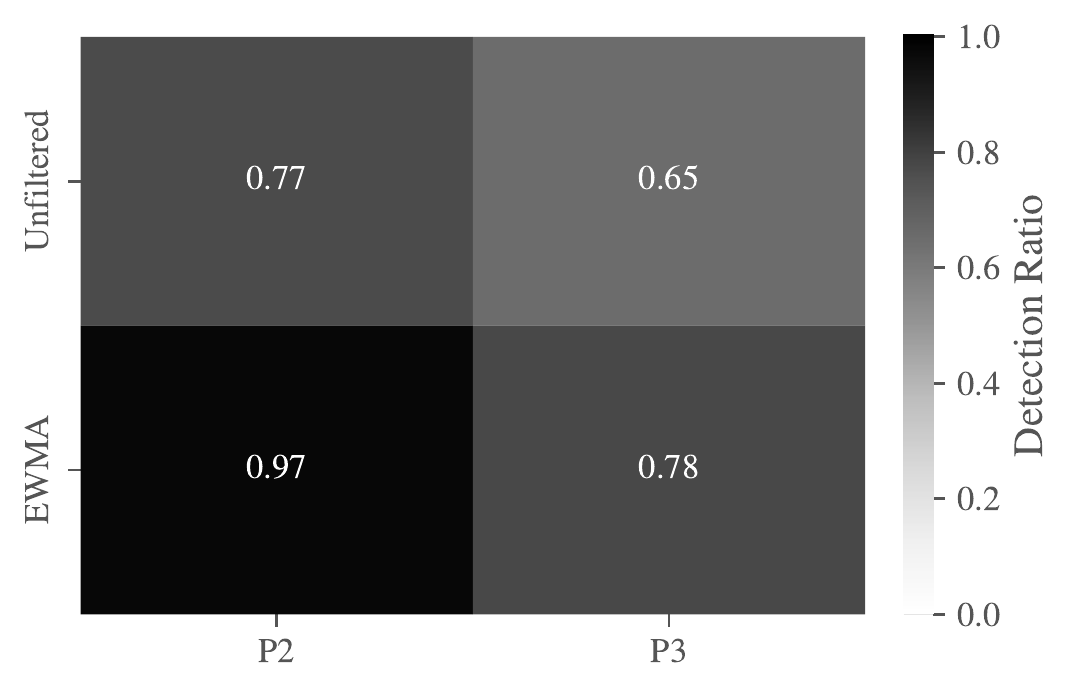}
    	\vspace{-0.06cm}
    	\caption{}
    	\label{fig:Matrix}
	\end{subfigure}
	\hspace{0.2cm}
	\begin{subfigure}[b]{0.48\textwidth}
		\centering
    	\includegraphics[trim=8 0 0 0, clip, width=1\linewidth]{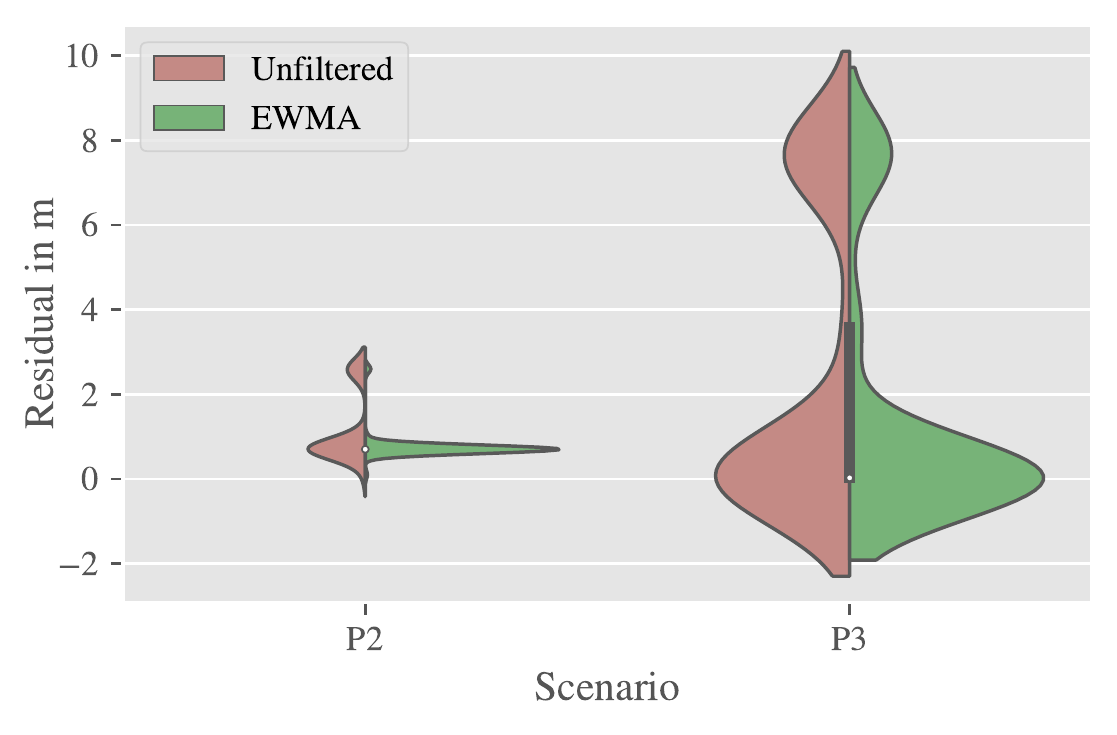}
    	\caption{}
    	\label{fig:Violin}
	\end{subfigure}%
	\centering
	\caption{Detection: \textbf{(a)} Detection Ratio Matrix; \textbf{(b)} Violin Plot: Unfiltered reflection path detection (red) and detection after EWMA filter of the CIR (green).}
	\label{fig:Detection}
\end{figure}

The mapping of the estimated reflection path with the elliptical model is shown in Fig.~\ref{fig:1_Lageplan} alongside with the interpolated heatmap based on all values from the subtracted CIR in Fig.~\ref{fig:1_Heatmap}. It visualizes the detection approach with the help of an interval of the estimated reflection path length for every parking lot. In this case the detection indicated an occupancy of the parking lot P2. The empirical evaluation of the detection ratio over all measurement epochs is visualized in Fig.~\ref{fig:Matrix}. For the discussed scenario P2, the detection ratio is $\SI{97}{\percent}$ when applying the EWMA filter, compared to only $\SI{77}{\percent}$ solely based on the unfiltered CIR. This shows the smoothing and outlier removing effects of the EMWA (cf. Fig.~\ref{fig:Violin}), allowing a more robust occupancy detection.

\subsection{Minibus on P3}
The previously discussed scenario in Fig.~\ref{fig:sideways_far} (cf. Sec.~\ref{sec:Approach}) depicts the detection scenario with the minibus on parking lot P3. For this scenario the reference reflection path length is \SI{11.84}{\meter}. From the corresponding CIRs in Fig.~\ref{fig:Subtraction} a reflection path length of \SI{11.93}{\meter} can be estimated. 

The background subtraction here eliminates a strong reflection, potentially originating from another vehicle in the environment and therefore allows the focus on the target object on P3. Another effect can be observed from Fig.~\ref{fig:EWMA}, where at \SI{19}{\meter} a double reflection becomes visible. The signal reflected at the target object is not only received via a single reflection, but also via a second reflection. In this case the subtraction also estimates the double reflection for the vehicle detection. In order to address this problem the number of used sensors should be increased to improve the robustness of the detection.

The mapping and heatmap (Fig.~\ref{fig:sideways_far}) of the subtracted CIR show a clear reflection at the minibus on P3. However, the detection rate (Fig.~\ref{fig:Matrix}) with $\SI{78}{\percent}$ is lower when compared to the P2 scenario.  This is also supported by Fig.~\ref{fig:Violin}, which depicts the residuals between the reference reflection lengths and the estimated lengths for both scenarios. For this, the aforementioned ambiguity for the reflection path in this scenario is revealed.

\subsection{Resolution and Shadowing}
\begin{figure}[pos=ht]
	\centering
	\begin{subfigure}[b]{0.34\textwidth}
		\centering
		\includegraphics[width=1\linewidth]{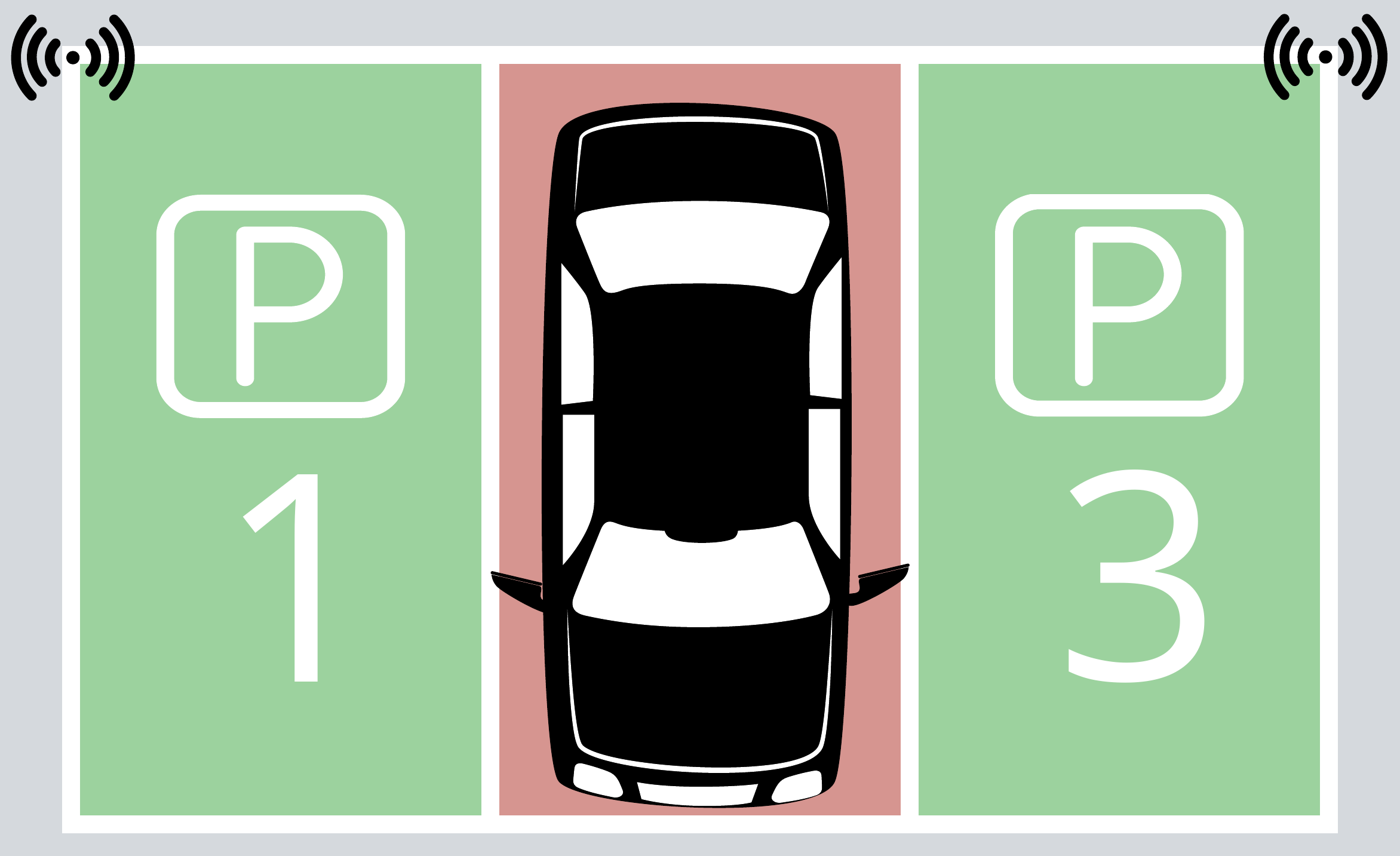}
		\vspace{-0.05cm}
		\caption{}
		\label{fig:4_Lageplan}
	\end{subfigure}
	\hspace{0cm}
	\begin{subfigure}[b]{0.31\textwidth}
		\centering
		\includegraphics[width=1\linewidth]{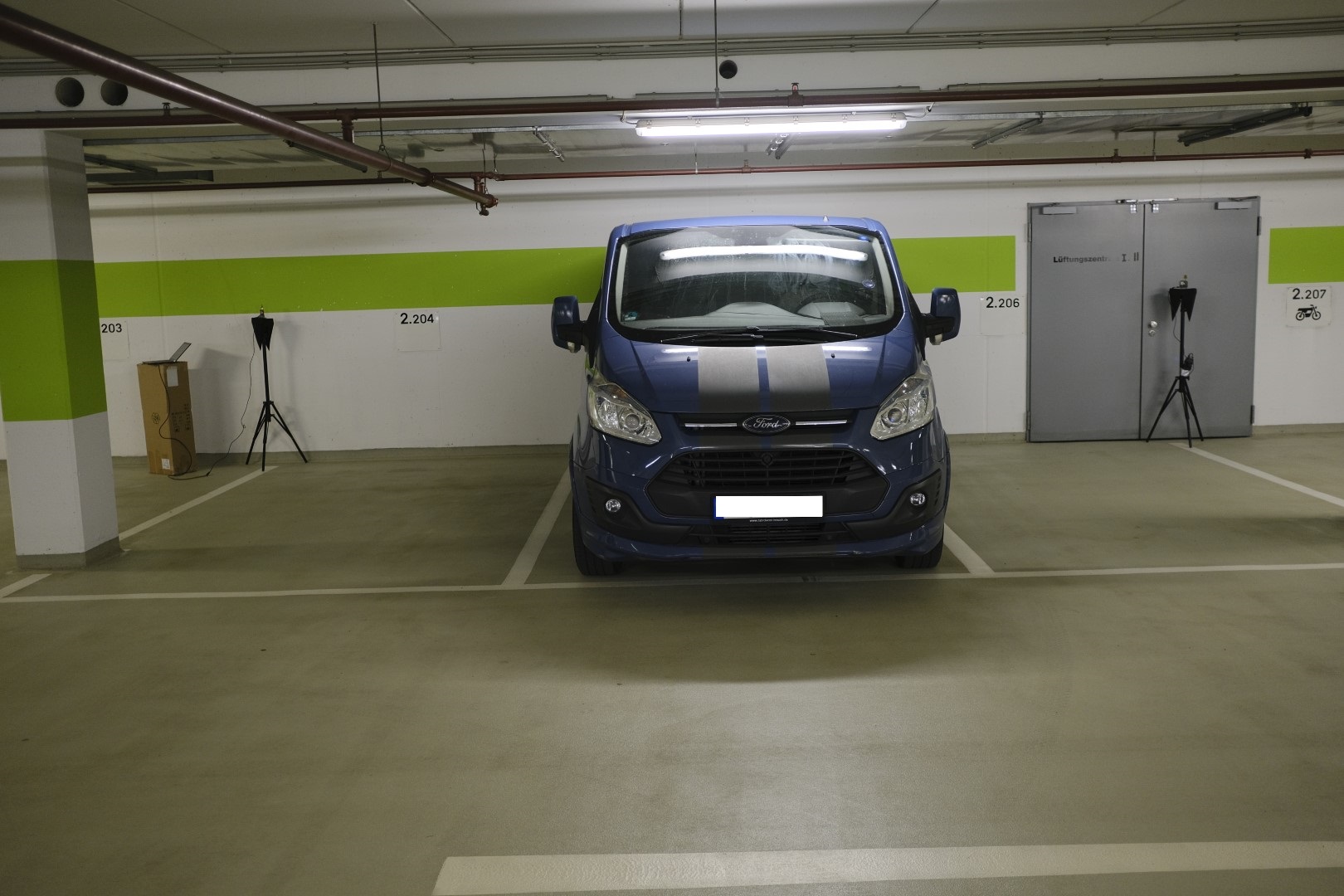}
		\vspace{-0.05cm}
		\caption{}
		\label{fig:4_Foto}
	\end{subfigure}%
	\centering
	\hspace{0cm}
	\begin{subfigure}[b]{0.32\textwidth}
		\centering
		\includegraphics[trim=0 10 0 0, clip, width=1\linewidth]{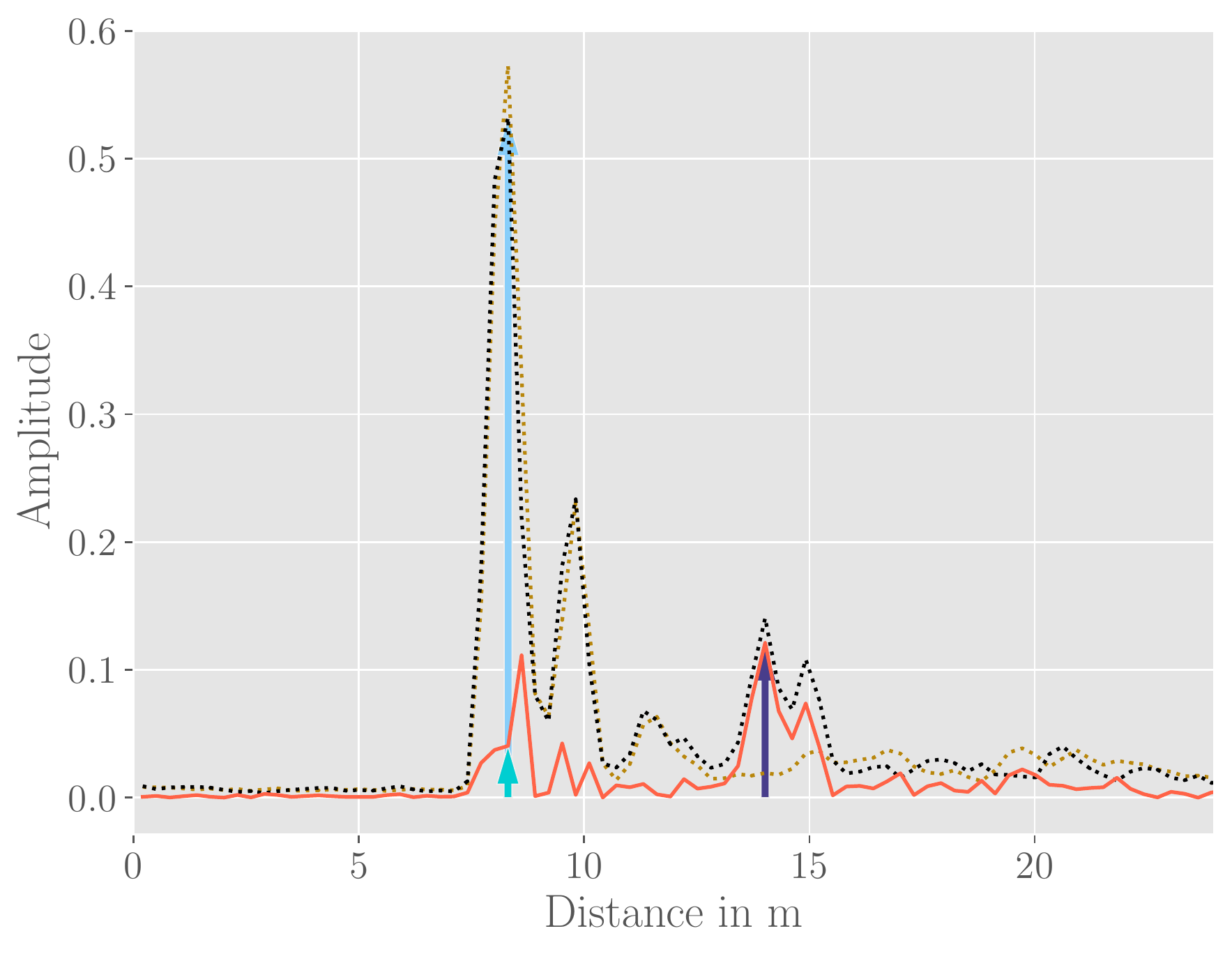}
		\caption{}
		\label{fig:4_CIR}
	\end{subfigure}
	\centering
	\caption{Scenario with transceivers along the wall: \textbf{(a)} Site plan; \textbf{(b)} Environment with vehicle; \textbf{(c)} CIRs with subtraction.}
	\label{fig:wand_unten}
\end{figure}

In order to further validate the occupancy detection, an additional transceiver arrangement is further examined. In these scenarios, the UWB modules are placed to cover the same three parking lots, but now along the wall, separated by \SI{8.25}{\meter} (cf. Fig.~\ref{fig:4_Lageplan}). Again, the calibration measurements were carried out without a vehicle and afterwards the minibus was parked on P2. This arrangement with the transceivers mounted at the walls is also more in line with a WSN for active localization. Additionally, it theoretically provides the capability to cover multiple parking lots and hence higher the coverage of the occupancy detection approach.

Furthermore, the CIR in Fig.~\ref{fig:4_CIR} indicates some limitations of the conducted measurement setup. Firstly, the separation of the direct and the reflection path in the CIR is examined. In this scenario the reference reflection path length is \SI{8.4}{\meter}, which is only \SI{0.15}{\meter} longer than the direct path. This path difference is not resolvable in the CIR with a resolution of \SI{0.3}{\meter}. The presences of the vehicle only results in a slightly wider direct path peak in the CIR, which is visible after the subtraction. From this it can be deduced that a minimum difference in the path lengths is required in order to separate them in the CIR. This is due to the bandwidth restrictions of the applied technology. For future transmission systems with higher available bandwidth, this shortcoming can potentially be compensated.

Another arising issue might be additional reflections, which can only be measured without the vehicle, but are blocked in the presence of it. This can be observed in Fig.~\ref{fig:4_CIR} (blue), where a reflection path lengths of \SI{13}{\meter} is estimated. A similar effect occurs in the field of camera-based object detection approaches, where shadowing within the images disturb the detection \cite{shadowing}.

\section{Conclusion \& Future Work}
\label{sec:Conclusion}
The rapidly increasing urbanization and associated demands on mobility and logistics are a challenging task for the upcoming years. With the help of digitization and networking, innovative, low-cost, and retrofittable systems can be established to tackle these tasks. Joint communication, localization, and sensing based on wireless transmission systems provide a promising, efficient, and innovative solution approach for these application fields, e.g.\ smart cities and parking. Therefore, this contribution presented a remote sensing approach based on multipath information to provide a passive occupancy detection for spatial resources like parking areas or mini-hubs. 

The proposed method uses CIR multipath data provided by the transmission modules in order to identify reflections within the propagation environment. To do so, we applied an EMWA filtering and background subtraction method to smooth the CIR and remove invariant components of it. The resulting CIR is then mapped to the environment using an elliptical model and an interpolated heatmap. Based on the reflection estimation, an occupancy detection for individual parking lots is derived.

Comparable occupancy detection technologies and methods (infrared, ultrasonic, camera) are dedicated for a single parking lot. With our setup we were able to observe recognizability of reflection with approximate \SI{25}{\meter} path length. For the coverage of an entire parking garage a comparably high number of devices would be required. However, the installed cooperative sensors serve a multi use purposes (joint communication, active localization, passive sensing) and therefore are more cost-efficient when compared to state-of-the-art occupancy detection systems.

For an application-close validation, preliminary field tests with a simplified measurement setup were conducted. The results revealed a proof-of-concept recognition of parking objects for limited constellations and therefore allowed a reliable occupancy detection rate. However, we also discussed the shortcomings of the used UWB technology as well as shadowing effects.

Further research will focus on an extended measurement campaign, including a network of UWB transceivers, allowing ambiguities to be resolved and providing redundant measurements to increase the robustness of the method. Also, detection of multiple objects in a wider range of scenarios will also be considered. Additionally, the use of beam-steering antenna for a better distinction between reflections is proposed.

\begin{figure}[pos=ht]
    \centering
    \includegraphics[width=1\linewidth]{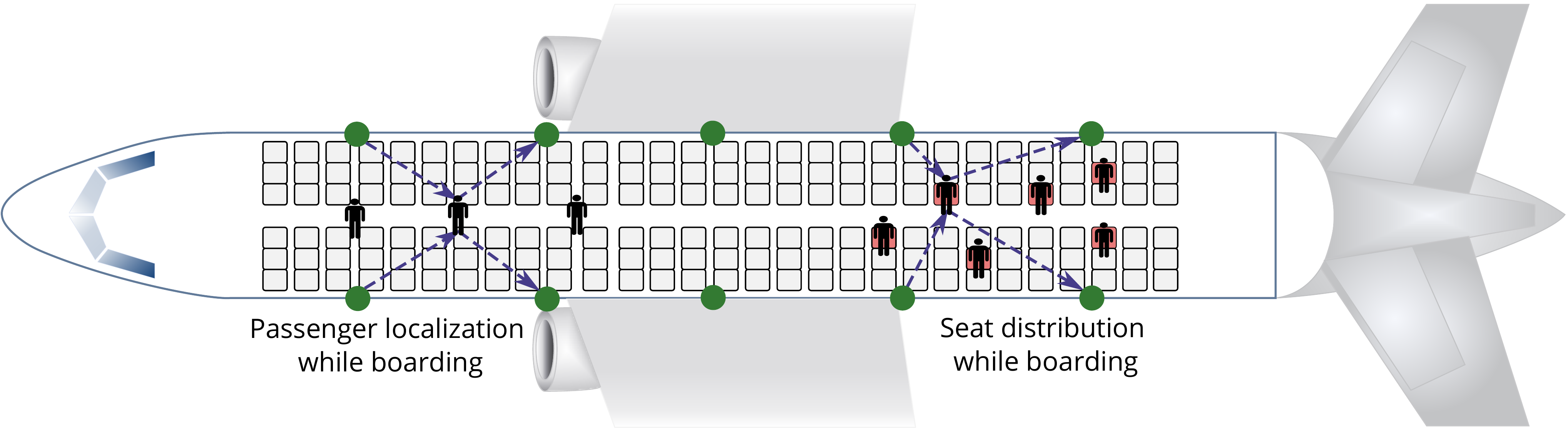}
    \caption{Future localization and detection radar system (green) in the connected aircraft cabin (Airbus A320) with passenger localization and seat distribution while boarding.}
    \label{fig:A320}
\end{figure}

In the future, a transfer of the approach to other physical layers, with advantageous properties, such as 5G/6G or Li-Fi, will be examined. In addition, validation could be extended to other environments and applications in ITS like the connected aircraft cabin \cite{covid-boarding}, where passive occupancy detection can be beneficial. Figure \ref{fig:A320} proposes a future cabin localization system based on the same multipath-assisted detection of people while boarding. This solution monitors the entire boarding process using a single WSN with communication \cite{Engelbrecht}, active localization \cite{Klipphahn} and radio sensing/passive detection.

\section{Acknowledgments}

This work has been supported by the European Union and the State of Saxony within the project "IVS-AMP" as well as by the Federal Ministry for Economic Affairs and Energy of Germany within the project "CAbiNET".

\begin{figure}[pos=h]
\centering
\includegraphics[width=0.4\textwidth]{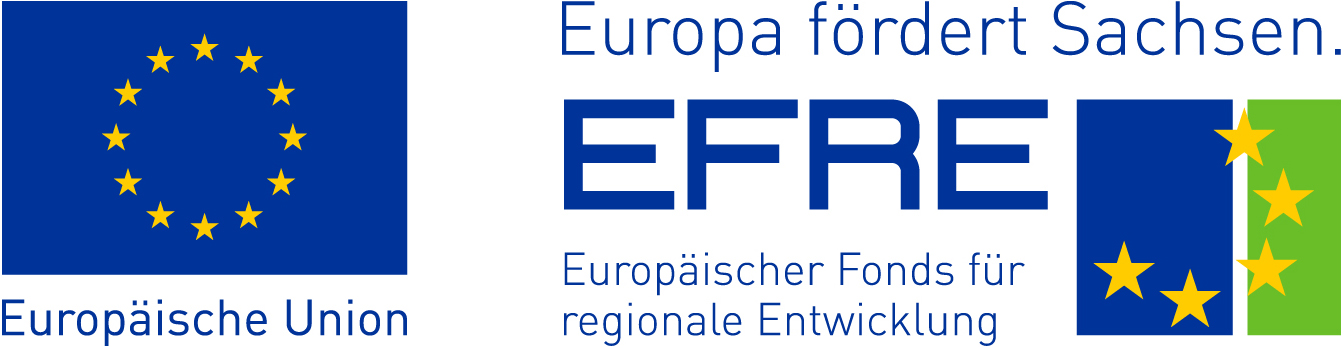}
\hspace{1cm}
\includegraphics[width=0.15\textwidth]{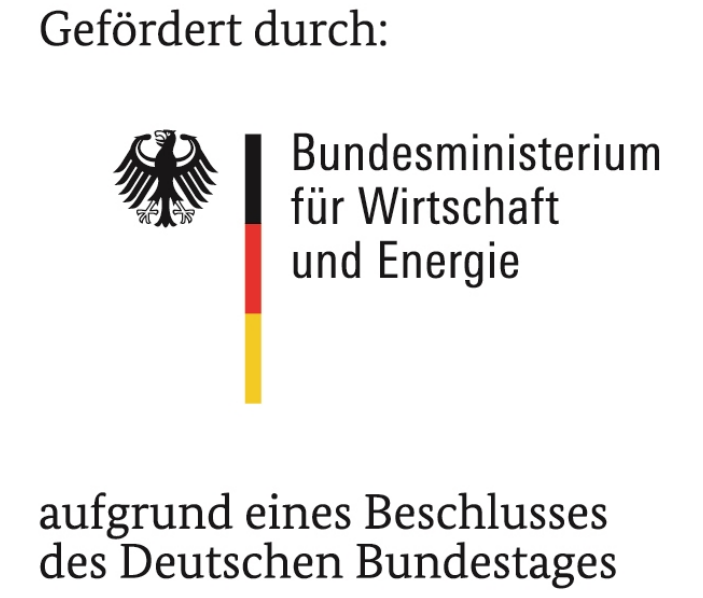}
\end{figure}
\vspace{0.5cm}

\bibliography{main}

\end{document}